\documentclass[,twocolumn]{aastex701}

\usepackage{amsmath}
\usepackage{multirow}

\begin{document}

\title{XClass: An Automated Multiwavelength Machine-Learning Pipeline for Classification of Extragalactic X-ray Sources. I. Pipeline Description}

\author[0000-0002-9282-5207]{Blagoy Rangelov}
\affiliation{Department of Physics, Texas State University, 601 University Drive, San Marcos, TX 78666, USA}
\email[hide]{rangelov@txstate.edu}  

\author{Nicholas Moore}
\affiliation{Department of Physics, Texas State University, 601 University Drive, San Marcos, TX 78666, USA}
\email[hide]{ZeroNMoore@outlook.com}  

\author{Haley Dutton}
\affiliation{Department of Physics, Texas State University, 601 University Drive, San Marcos, TX 78666, USA}
\email[hide]{hmd66@txstate.edu}

\author[0009-0000-5037-5458]{Erika Marentes}
\affiliation{Department of Physics, Texas State University, 601 University Drive, San Marcos, TX 78666, USA}
\email[hide]{e\_m474@txstate.edu}


\begin{abstract}
Most X-ray sources detected in nearby galaxies by the \textit{Chandra X-ray Observatory} lack astrophysical classifications. We present XClass (X-ray Classifier for extragalactic sources), an end-to-end machine-learning pipeline that classifies extragalactic X-ray point sources into seven classes: AGN, LMXBs, HMXBs, CVs, low-mass and high-mass foreground stars, and supernova remnants. The key challenge is photometric heterogeneity: training sources are predominantly Galactic with photometry from wide-field surveys (PanSTARRS and 2MASS), while extragalactic targets require \textit{Hubble Space Telescope} (HST) imaging in a disjoint filter system. We address this through a spectral energy distribution (SED) translation that fits class-appropriate spectral models to each training source and convolves the best-fit model through HST filter curves, producing synthetic magnitudes in a common feature space. The classifier uses an asymmetric two-stage Random Forest: Stage~1 separates broad categories (AGN, X-ray binaries, SNRs, stars) and Stage~2 resolves X-ray binaries into LMXBs and HMXBs using an augmented feature vector that includes Stage~1 probabilities. The training set is assembled from ten Galactic catalogs and extragalactic SNR catalogs, cross-matched with the Chandra Source Catalog v2.1. Features include X-ray hardness ratios, SED-translated HST colors, and X-ray-to-optical flux ratios. We restrict the training set to sources with at least one optical magnitude, avoiding imputation artifacts; the pipeline achieves 99.6\% accuracy and balanced accuracy of 0.90 on the resulting 11{,}374-source optical baseline, with excellent calibration (ECE = 0.002). XClass is modular, generalizable to any HST filter configuration, and will be applied to M31 and M33 in a companion paper.
\end{abstract}



\section{Introduction}\label{sec:intro}

A census of extragalactic X-ray source populations is essential for addressing questions ranging from X-ray binary evolution to the star formation histories of host galaxies \citep[e.g.,][]{Fabbiano2006, Gilfanov2004}. The \textit{Chandra X-ray Observatory} has enabled the detection of thousands of discrete X-ray sources in galaxies within $\sim$10~Mpc (e.g., \citealt{Evans2010,Evans2020}), yet the vast majority remain unclassified. In M31 and M33 alone, $\sim$3{,}000 X-ray sources have been cataloged, but fewer than $\sim$1/4 have reliable astrophysical classifications \citep[e.g.,][]{Stiele2011, Williams2018, Tullmann2011}.

In the absence of spectra, classification must rely on broadband photometric and X-ray properties --- hardness ratios, colors, and X-ray--to--optical flux ratios \citep[e.g.,][]{Kaplan2006, Misanovic2010, Brassington2012}. However, traditional two-parameter diagnostic approaches become impractical for thousands of sources in a high-dimensional feature space and do not provide quantified classification confidences. Machine-learning (ML) methods such as Random Forests \citep{Breiman2001} offer a solution, since they operate on many features simultaneously and provide probabilistic class assignments. Several groups have applied ML to X-ray source classification, primarily in Galactic fields \citep{Lo2014, Tranin2022}. \citet{Yang2022} presented MUWCLASS, a Random Forest pipeline using X-ray, optical, NIR, and IR photometry from the Chandra Source Catalog (CSC), PanSTARRS, \textit{Gaia}, 2MASS, and \textit{WISE}, which was used to classify $\sim$66{,}000 CSC~v2.1 sources and subsequently applied to NGC~3532 \citep{Chen2023} and to 13 unidentified \textit{Fermi} LAT sources \citep{Rangelov2024}. In the extragalactic domain, \citet{Arnason2020} used Random Forests with X-ray properties alone to perform binary XRB/non-XRB classification of 943 \textit{Chandra} sources in M31; however, that work did not incorporate optical or NIR photometry.

All of the above efforts share a common limitation: the training and application data use the same photometric system. Extending ML classification to extragalactic sources using multiwavelength (MW) data introduces a fundamental challenge --- photometric heterogeneity between the training and application domains. Confidently classified training sources are overwhelmingly Galactic, with counterparts characterized by wide-field surveys (PanSTARRS and 2MASS), while extragalactic targets require \textit{Hubble Space Telescope} (HST) imaging to resolve faint counterparts in crowded fields. Given that the HST filter system is entirely disjoint from the ground-based sets, direct feature-space comparison is not possible.

In this paper, we present \textbf{XClass} (\textbf{X}-ray \textbf{Class}ifier for extragalactic sources), an end-to-end ML pipeline for classifying extragalactic X-ray point sources into seven classes: active galactic nuclei (AGN), low mass X-ray binaries (LMXBs), high mass X-ray binaries (HMXBs), cataclsysmic variables (CVs), low-mass foreground stars (LM-STARs), high-mass foreground stars (HM-STARs), and supernova remnants (SNRs). XClass builds on the conceptual framework of MUWCLASS but is redesigned for the extragalactic context. The key innovation is a \emph{spectral energy distribution (SED) translation} procedure: for each training source, a class-appropriate physical SED model is fit to the ground-based photometry and convolved through HST filter curves, producing  synthetic HST magnitudes that place training and application data into a common feature space. The approach is general, accommodating any combination of input surveys and output HST filters.

In addition to the SED translation, XClass incorporates: (1)~an asymmetric two-stage Random Forest that separates broad categories before resolving the LMXB/HMXB distinction; (2)~a training dataset assembled from ten literature catalogs of Galactic sources, augmented with extragalactic SNR catalogs; (3)~cross-matching with CSC~2.1 and photometric queries to PanSTARRS~DR2 and 2MASS; and (4)~counterpart matching with the Hubble Source Catalog v3 for target galaxies. To our knowledge, XClass is among the first ML pipelines designed specifically for extragalactic X-ray source classification using MW data across heterogeneous filter systems. Given the growing \textit{Chandra} and HST archives --- and the advent of \textit{Athena} \citep{Nandra2013}, AXIS \citep{Mushotzky2018}, Lynx \citep{Gaskin2019}, the Einstein Probe \citep{Yuan2022}, the \textit{Vera C.~Rubin Observatory} \citep{Ivezic2019}, and the \textit{Roman Space Telescope} --- such a tool is both timely and necessary. 

The rest of the paper is organized as follows. Section~\ref{sec:training_data} describes the training dataset. Section~\ref{sec:sed_translation} presents the SED translation procedure. Section~\ref{sec:classification} covers feature engineering and the two-stage classifier. Section~\ref{sec:performance} evaluates classification performance. We discuss design decisions and limitations in Section~\ref{sec:discussion} and summarize in Section~\ref{sec:summary}.

\begin{figure*}
\centering
\includegraphics[width=\textwidth]{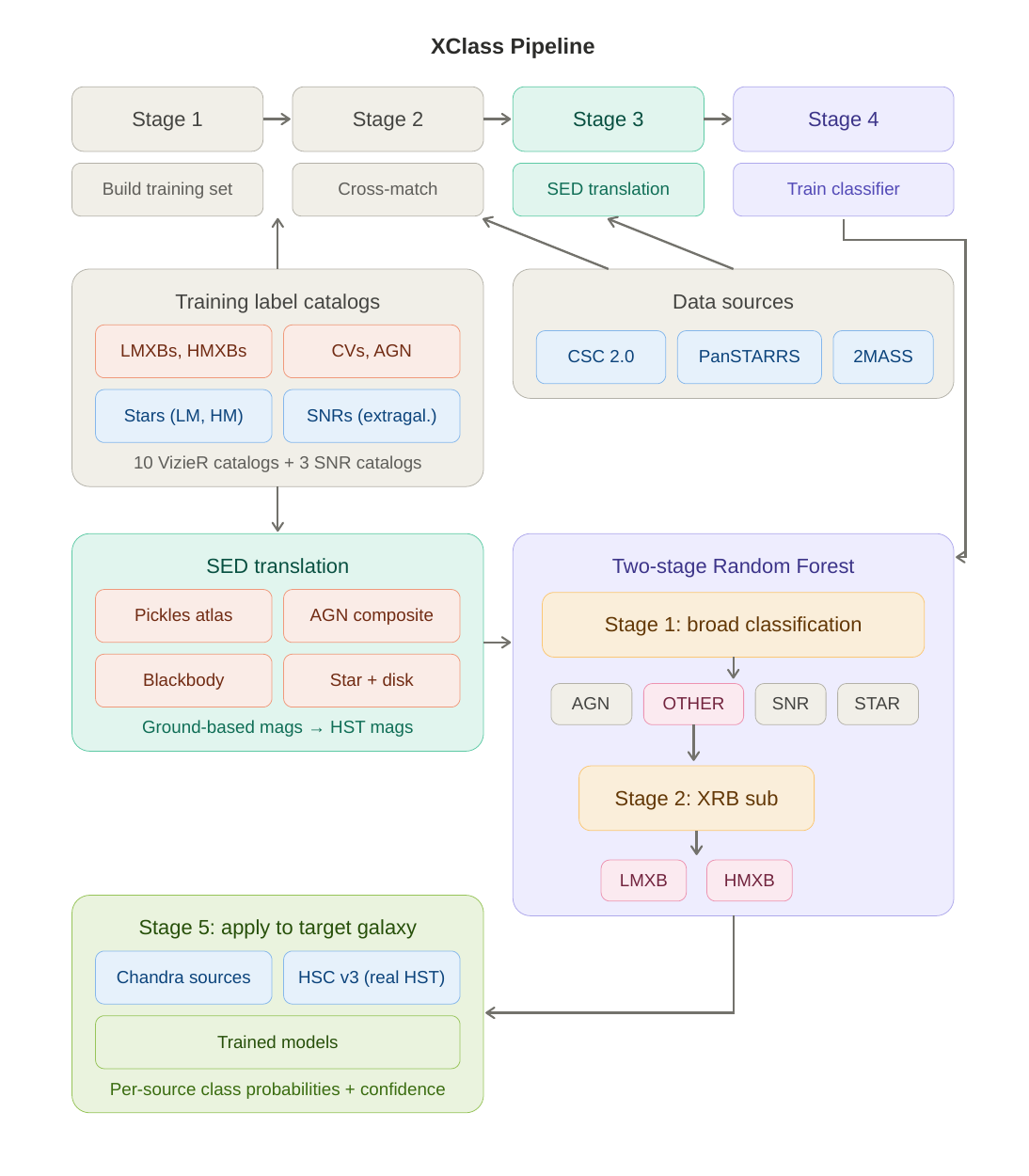}
\caption{End-to-end schematic of the XClass pipeline. The five stages are shown from left to right: (1)~training-set construction from literature catalogs, (2)~cross-matching with the Chandra Source Catalog and ground-based photometric surveys, (3)~SED translation from ground-based to HST photometric features (Section~\ref{sec:sed_translation}), (4)~two-stage Random Forest training (Section~\ref{sec:classification}), and (5)~application to target galaxies using real HST photometry from the Hubble Source Catalog. The two-stage classifier architecture is shown as an inset: Stage~1 classifies sources into four broad groups (AGN, OTHER, SNR, STAR), and Stage~2 resolves 
OTHER into LMXB or HMXB.}
\label{fig:pipeline}
\end{figure*}


\section{Training Data}\label{sec:training_data}

The XClass training dataset is assembled from literature catalogs of confidently classified X-ray sources, cross-matched with detections in the Chandra Source Catalog, and augmented with ground-based MW photometry.

\subsection{Source Classes and Label Catalogs}\label{sec:classes}

XClass classifies sources into seven astrophysical classes, selected to represent the dominant X-ray point-source populations in nearby spiral galaxies. The class scheme builds on the Galactic MUWCLASS pipeline \citep{Yang2022}, omitting isolated neutron stars (too faint at extragalactic distances) and young stellar objects (unresolved in external galaxies), and absorbing Wolf-Rayet stars into the high-mass star class. Training labels were drawn from ten literature catalogs queried from VizieR \citep{Ochsenbein2000} via \texttt{astroquery} \citep{Ginsburg2019}, plus three extragalactic SNR 
catalogs.

\textbf{AGN} --- background sources seen in projection or nuclear sources within the target galaxy --- were drawn from the V\'{e}ron-Cetty \& V\'{e}ron catalog of quasars and active nuclei \citep{VeronCetty2010}. AGN are the most numerous class in both the training set and typical extragalactic X-ray catalogs \citep{Stiele2011, Williams2018}.

\textbf{LMXBs}, containing a compact object accreting from a low-mass ($\lesssim 1\,M_\odot$) donor via Roche lobe overflow, trace the old stellar population \citep{Gilfanov2004}. Labels were compiled from the Milky Way LMXB catalog of \citet{Liu2007} and the X-ray binary tables of \citet{RitterKolb2003}.

\textbf{HMXBs}, with compact objects accreting from massive ($\gtrsim 8\,M_\odot$) OB or Be-star donors, are associated with recent star formation \citep{Grimm2003}. Labels were taken from the HMXB catalog of \citet{Liu2006}.

\textbf{CVs} are white-dwarf binaries accreting from low-mass companions. Although intrinsically faint in X-rays, they appear as foreground contaminants. Labels were drawn from the CV tables of \citet{RitterKolb2003} and the atlas of \citet{Downes2001}.

\textbf{LM-STARs} of spectral types A--M emit X-rays through coronal activity and are ubiquitous foreground contaminants.  \textbf{HM-STARs} of spectral types O--B (including Wolf-Rayet stars) produce X-rays via stellar winds. Stellar labels were obtained primarily from the MK spectral classification catalog of \citet{Skiff2014}, with O/B types assigned to HM-STAR and later types to LM-STAR; ambiguous types were excluded. These were supplemented by APOGEE-2 DR16 \citep{Majewski2017} sources (with valid $T_\mathrm{eff}$, $\log g$, and radial-velocity scatter $\leq 1.0$~km~s$^{-1}$) as LM-STARs, and by Wolf-Rayet stars from \citet{vanderHucht2001, vanderHucht2006} as HM-STARs.

\textbf{SNRs} are the sole class trained on extragalactic data, since Galactic SNRs observed by \textit{Chandra} are typically highly extended and poorly representative of the compact remnants seen in external galaxies. We use SNR catalogs from M33 \citep{Long2010}, NGC~6946 \citep{Long2019}, and M83 \citep{Blair2012}, cross-matched to \textit{Chandra} detections at a fixed radius of $3\farcs0$. SNRs are excluded from the SED translation step (Section~\ref{sec:sed_translation}) because their training features rely only on HST photometry when available, or on X-ray properties alone.

\begin{deluxetable}{lccccc}
\tablecaption{Training dataset summary by class. Columns show the number of sources at each pipeline stage: (1)~after catalog compilation and deduplication, (2)~after cross-matching with CSC~2.1 ($\geq 3\sigma$ detections), (3)~after SED translation (sources with $\geq 3$ photometric 
bands), and (4)~final count entering the classifier, restricted to sources 
with $\geq 1$ non-NaN predicted HST magnitude (see 
Section~\ref{sec:splitting}).\label{tab:td_summary}}
\tablehead{
\colhead{Class} & \colhead{Catalogs$^a$} & \colhead{CSC Match} & 
\colhead{SED ($\geq$3)} & \colhead{Classifier$^b$}
}
\startdata
AGN      & 168{,}314 & 5{,}954 & 5{,}231 & 5{,}231 \\
LMXB     &       255 &     100 &      82 &      86 \\
HMXB     &       114 &      37 &      36 &      39 \\
CV       &   1{,}429 &      98 &      87 &      86 \\
LM-STAR  & 947{,}698 &  4{,}693 & 4{,}615 & 4{,}610 \\
HM-STAR  &  58{,}085 &  1{,}232 & 1{,}215 & 1{,}214 \\
SNR      &       639 & \nodata &    \nodata &     108$^c$ \\
\hline
Total    & 1{,}176{,}534 & 12{,}114 & 11{,}266 & 11{,}374 \\
\enddata
\tablenotetext{a}{Counts before any X-ray cross-matching; the large stellar totals reflect the full Skiff and APOGEE catalogs prior to positional matching with Chandra.}
\tablenotetext{b}{Sources with $\geq 1$ non-NaN predicted HST magnitude from SED translation or, for SNRs, from direct HSC photometry. Sources with zero optical features are excluded to avoid median-imputation artifacts (Section~\ref{sec:splitting}).}
\tablenotetext{c}{SNRs bypass SED translation; the 108 sources reflect those with $\geq 1$ PHAT-filter detection in the Hubble Source Catalog.}
\end{deluxetable}

\subsection{Deduplication}\label{sec:dedup}

After concatenating all catalog entries, deduplication was applied within each source class using a Union-Find data structure with two merging criteria: (1)~rows with identical normalized names were merged unconditionally, and (2)~positional pairs within $5\arcsec$ were merged if they share a name, either name is empty, or their separation is $\leq 1\arcsec$. For each merged group, the canonical position is the median R.A.\ and decl., and the canonical name is the shortest non-empty string. SNR deduplication uses a tighter $2\arcsec$ radius with priority given to X-ray detections over optical or radio.

\subsection{Cross-matching with the Chandra Source Catalog}\label{sec:csc_match}

The deduplicated training-set positions were cross-matched against CSC~2.1 \citep{Evans2010, Evans2020}, which contains $\sim$317{,}000 unique X-ray sources with positions, fluxes in four energy bands (soft: 0.5--1.2~keV, medium: 1.2--2.0~keV, hard: 2.0--7.0~keV, broad: 0.5--7.0~keV), and positional uncertainties. The effective positional uncertainty for each Chandra source was computed from the error ellipse semi-axes $r_0$ and $r_1$ as $\sigma_\mathrm{eff} = \sqrt{(r_0^2 + r_1^2)/2}$. The search radius was set to $R_\mathrm{search} = \max(3\,\sigma_\mathrm{eff},\; 0\farcs5)$, and only sources with detection significance $\geq 3\sigma$ were included.

For each matched pair, we compute the normalized separation $\nu = \theta / \sigma_\mathrm{eff}$, where $\theta$ is the angular separation. Matches with $\nu \leq 2.0$ are flagged as ``secure,'' those with $2.0 < \nu \leq 3.0$ as ``ambiguous,'' and those with $\nu > 3.0$ are excluded. The normalized separation is retained and can be used as a classification feature (Section~\ref{sec:classification}). The X-ray flux uncertainty for each band is estimated as $\sigma_{F_x} = (F_{x,\mathrm{hi}} - F_{x,\mathrm{lo}})/2$.

\subsection{Ground-Based Photometry}\label{sec:photometry}

For each training source matched to a Chandra detection, MW photometry was queried from two surveys: PanSTARRS~DR2 \citep{Chambers2016} ($grizy$; Dec $> -30\degr$) and 2MASS \citep{Skrutskie2006} ($JHK_s$). The nearest counterpart within $5\arcsec$ was selected for each source. 

\subsection{Training Set Summary}\label{sec:td_summary}

\begin{figure}
\centering
\includegraphics[width=\columnwidth]{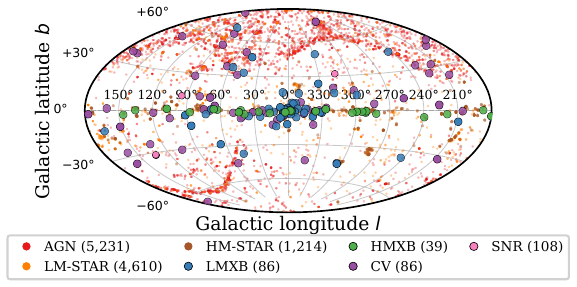}
\caption{Sky distribution of the 11{,}374 training sources in Galactic coordinates (Aitoff projection, Galactic center at center). Sources are colored by class; the dashed line marks the Galactic plane ($b = 0$). The training set is overwhelmingly Galactic: the accreting-binary and massive-star classes concentrate strongly toward the plane (median $|b| = 1.1^\circ$ for HMXBs, $1.2^\circ$ for HM-STARs, $3.7^\circ$ for LMXBs), while low-mass stars, being local, are more broadly spread (median $|b| = 18.8^\circ$). The AGN labels, drawn from optical quasar catalogs cross-matched to the CSC, are predominantly high-latitude (median $|b| = 52.7^\circ$). This reflects the selection function of optical AGN identification, which is suppressed at low latitude by Galactic extinction and crowding, rather than the intrinsic distribution of background AGN. The supernova remnants --- the only class trained on extragalactic data --- appear as compact clumps at their host galaxies M83, NGC~6946, and M33 (all within $2^\circ$ of one of the three). This distribution illustrates the predominantly Galactic character of the training set (Section~\ref{sec:limitations}).}
\label{fig:skymap}
\end{figure}

Table~\ref{tab:td_summary} summarizes the number of sources per class at each stage. We note that the training set exhibits substantial class imbalance, with AGN dominating and compact-object classes (LMXB, HMXB, CV) being underrepresented. This motivates the use of balanced class weighting during classifier training (Section~\ref{sec:classification}). Figure~\ref{fig:skymap} shows the sky distribution of the training sources in Galactic coordinates. The accreting-binary and massive-star classes concentrate toward the Galactic plane, while the supernova remnants --- the only extragalactic-trained class --- form compact groups at their host galaxies; this illustrates the predominantly Galactic character of the training set (Section~\ref{sec:limitations}).

\section{SED Translation}\label{sec:sed_translation}

The central methodological innovation of XClass is the SED translation step, which bridges the photometric gap between the ground-based surveys used to characterize training sources and the HST photometry available for extragalactic targets. Training sources have photometry from PanSTARRS ($grizy$) and 2MASS ($JHK_s$), while extragalactic targets are observed through entirely different HST filters (e.g., F275W, F336W, F475W, F814W, F110W, F160W for the PHAT survey; \citealt{Dalcanton2012}). Simple linear color transformations \citep[e.g.,][]{Sirianni2005} are inadequate because they are source-type dependent --- the correction between PanSTARRS $i$ and HST F814W differs for a power-law AGN, a K-type star, and a flat accretion disk --- thus introducing systematic biases for precisely the classes the classifier must distinguish. The SED translation resolves this by fitting a physically motivated spectral model to each training source and convolving the best-fit model through the HST filter curves, producing synthetic HST magnitudes that preserve class-discriminating color information.

\subsection{Physical SED Models and Class Assignment}\label{sec:sed_models}

Filter transmission curves for both input (ground-based) and output (HST) systems were obtained from the SVO Filter Profile Service \citep{Rodrigo2020}. The HST set covers $\sim$40 filters across ACS/WFC, WFC3/UVIS, and WFC3/IR. For a specific target galaxy, only the relevant subset can be used as classification features. All SED models are evaluated on a common wavelength grid spanning 100--25{,}000~\AA\ in 5~\AA\ steps. Four model families are employed.

\textbf{Pickles stellar atlas.} Empirical spectra from the \citet{Pickles1998} UV--optical--IR library, covering 27 spectral types from O5 to M6 (loaded from the STScI CDBS archive). For sources with a known spectral type, the nearest library type is selected; otherwise, all 27 templates are searched.

\textbf{AGN composite.} The mean SDSS quasar spectrum of \citet{VandenBerk2001}, covering the UV through NIR with the characteristic power-law continuum and broad emission lines.

\textbf{Power-law.} $f_\nu(\lambda) \propto (\lambda_\mathrm{ref}/\lambda)^\alpha$ with $\lambda_\mathrm{ref} = 5500$~\AA\ and $\alpha \in [-2.5, +1.5]$ in steps of 0.05 (81 grid points). This is used as a fallback for AGN.

\textbf{Blackbody.} A Planck function with a temperature grid spanning 2{,}000--9{,}750~K in 250~K steps and 10{,}000--50{,}000~K in 1{,}000~K steps (73 grid points). This is used as a fallback for stars and X-ray binaries.

\textbf{Two-component star + disk.} For X-ray binaries and CVs, the optical emission is modeled as
\begin{equation}\label{eq:twocomp}
    f_\nu(\lambda) = (1 - f_\mathrm{disk}) \cdot f_{\nu,\mathrm{star}}(\lambda) 
    + f_\mathrm{disk} \cdot f_{\nu,\mathrm{disk}}(\lambda),
\end{equation}
where $f_{\nu,\mathrm{star}}$ is a Pickles template (K5 for LMXBs/CVs, B2 for HMXBs; \citealt{Charles2006}) and $f_{\nu,\mathrm{disk}}$ is a flat ($f_\nu = \mathrm{const}$) spectrum representing a steady-state accretion disk. 
The disk fraction is varied over $f_\mathrm{disk} \in \{0.0, 0.1, 0.2, 0.3, 0.4, 0.5\}$.

Each class is assigned a primary and fallback model as summarized in Table~\ref{tab:sed_models}. SNRs are excluded from SED translation entirely (see Section~\ref{sec:classes}).

\begin{deluxetable}{lll}
\tablecaption{SED model assignment by source class. The primary model is used first; if it fails, the fallback model is attempted. SNRs bypass SED translation and use direct HST photometry when available.\label{tab:sed_models}}
\tablehead{
\colhead{Class} & \colhead{Primary Model} & \colhead{Fallback Model}
}
\startdata
AGN     & AGN composite       & Power-law \\
LMXB    & Two-component (K5 + disk) & Blackbody \\
HMXB    & Two-component (B2 + disk) & Blackbody \\
CV      & Two-component (K5 + disk) & Blackbody \\
LM-STAR & Pickles atlas       & Blackbody \\
HM-STAR & Pickles atlas       & Blackbody \\
\enddata
\end{deluxetable}

\subsection{Fitting Procedure}\label{sec:fitting}

The SED fit is applied independently to each training source with at least three valid photometric bands.

\textbf{Step~1: Photometric conversion.} Magnitudes were converted to flux densities $f_\nu$ (erg~s$^{-1}$~cm$^{-2}$~Hz$^{-1}$) using the AB zero point (3631~Jy) for PanSTARRS, and Vega zero points for 2MASS ($Z_\nu = 1594$, 1024, 666.7~Jy for $J$, $H$, $K_s$). Uncertainties were propagated as $\sigma_{f_\nu} = f_\nu \cdot (\ln 10 / 2.5) \cdot \sigma_m$; when unavailable, 10\% of the flux was adopted.

\textbf{Step~2: Normalization.} The band with the highest observed $f_\nu$ was selected as the normalization anchor.

\textbf{Step~3: Grid search.} For each model parameter (temperature, spectral index, disk fraction, or spectral type), the model was normalized to match the anchor-band flux and the reduced chi-squared was evaluated as
\begin{equation}\label{eq:chisq}
    \chi^2_\mathrm{red} = \frac{1}{N_\mathrm{bands} - 1} 
    \sum_{k \neq \mathrm{norm}} 
    \left(\frac{f_{\nu,k}^\mathrm{obs} - \langle f_\nu \rangle_k^\mathrm{model}}
    {\sigma_{f_{\nu,k}}}\right)^2,
\end{equation}
where the synthetic flux through filter $k$ is calculated as
\begin{equation}\label{eq:synth_flux}
    \langle f_\nu \rangle_k = \frac{\int f_\nu(\lambda)\, T_k(\lambda)\, 
    d\lambda}{\int T_k(\lambda)\, d\lambda}.
\end{equation}

\textbf{Step~4: Best model selection.} The grid point minimizing $\chi^2_\mathrm{red}$ was selected. If the primary model fails, the fallback is attempted; sources for which both fail are excluded.

\textbf{Step~5: HST magnitude prediction.} The best-fit SED was convolved through each target HST filter via Equation~(\ref{eq:synth_flux}), yielding predicted AB magnitudes $m_\mathrm{pred} = -2.5\,\log_{10}(\langle f_\nu \rangle) - 48.60$. These translated magnitudes place training sources into the same feature space as the HST-observed extragalactic application data.

\subsection{Uncertainty Propagation}\label{sec:sed_uncertainty}

The uncertainty on each predicted magnitude is estimated as
\begin{equation}\label{eq:base_err}
    \sigma_\mathrm{base} = \min\!\left(0.1 \cdot \sqrt{\chi^2_\mathrm{red}},\ 
    1.0~\mathrm{mag}\right).
\end{equation}
For UV filters (F275W, F336W), an additional 0.4~mag systematic is added in quadrature:
\begin{equation}\label{eq:uv_err}
    \sigma_\mathrm{UV} = \sqrt{\sigma_\mathrm{base}^2 + 
    (0.4~\mathrm{mag})^2}.
\end{equation}
This reflects the absence of ground-based constraints shortward of $\sim$4800~\AA\ (PanSTARRS $g$-band). Thus, UV predictions are model extrapolations rather than interpolations. We note that the SED translation is general: applying XClass to a galaxy observed with a different HST filter set requires only specifying the output filter names, and incomplete photometric coverage is handled naturally through degraded fit quality propagated into the feature uncertainties.

\subsection{SED Fit Quality}\label{sec:sed_quality}

\begin{figure}
\centering
\includegraphics[width=\columnwidth]{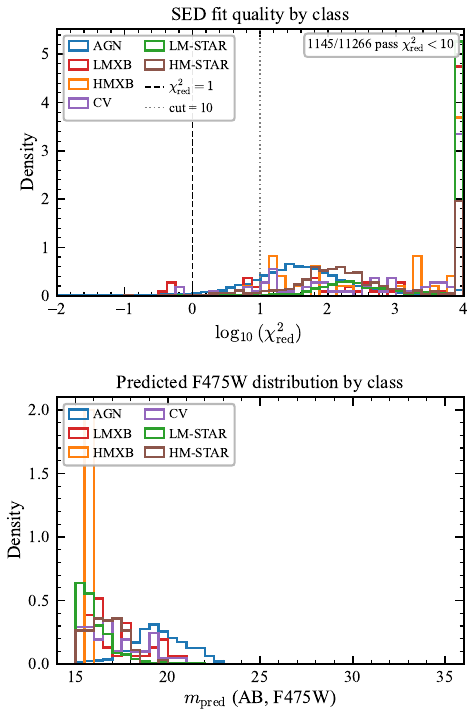}
\caption{SED fit quality diagnostics. \textit{Top:} Distribution of $\log_{10}(\chi^2_\mathrm{red})$ for all non-SNR training sources, color-coded by class; the dashed line marks $\chi^2_\mathrm{red} = 10$.  \textit{Bottom:} Distribution of predicted F475W magnitudes by class.}
\label{fig:sed_quality}
\end{figure}

The quality of the SED fits varies across the training set. Figure~\ref{fig:sed_quality} shows the distribution of $\chi^2_\mathrm{red}$ values for all non-SNR training sources, broken down by class. The median $\chi^2_\mathrm{red}$ is $\sim$177, with only $\sim$10\% of sources achieving $\chi^2_\mathrm{red} < 10$. However, we note that a high $\chi^2_\mathrm{red}$ does not necessarily indicate an inadequate SED model; it can equally reflect noisy input photometry, underestimated photometric uncertainties (a default of 10\% is adopted when per-band errors are unavailable), or incorrect cross-matches in crowded fields. All three effects inflate $\chi^2_\mathrm{red}$ without implying that the predicted colors are physically unreasonable.

For the subset of sources with well-constrained fits ($\chi^2_\mathrm{red} < 10$), we verified the internal consistency of the synthetic photometry by comparing predicted HST magnitudes with the input PanSTARRS photometry in the overlapping wavelength range. The predicted F475W and F814W magnitudes reproduce the input PS1 $g$ and $i$ photometry with median offsets of $+0.03$ and $-0.07$~mag and scatter (MAD) of 0.05 and 0.06~mag, respectively. The small systematic offsets are consistent with the $\sim$100~\AA\ wavelength difference between the PS1 and ACS filter pivot wavelengths, confirming that the synthetic photometry machinery is numerically accurate.

To investigate the origin of the high $\chi^2_\mathrm{red}$ values, we performed an ablation using only PanSTARRS $grizy$ photometry as SED input (excluding the 2MASS NIR bands). The overall median $\chi^2_\mathrm{red}$ decreased modestly (177.6 to 92.1), but the improvement was class-dependent: HM-STARs improved substantially (median $\chi^2_\mathrm{red}$ from 209 to 60), while X-ray binary fits degraded without the NIR anchor needed to constrain the disk fraction. AGN $\chi^2_\mathrm{red}$ was unchanged (37.2 vs.\ 36.5), indicating that intrinsic SED diversity --- not inter-survey calibration --- is the dominant source of fit residuals. The degradation of the X-ray binary fits without the NIR bands confirms that 2MASS photometry is important for constraining the two-component (donor + disk) SED model. We therefore retain the full multi-survey input as the default configuration.

For sources with higher $\chi^2_\mathrm{red}$, the uncertainty propagation (Equation~\ref{eq:base_err}) automatically assigns larger predicted-magnitude errors, appropriately down-weighting the optical/NIR color features for poorly constrained sources. The cross-validation performance reported in Section~\ref{sec:performance} represents the end-to-end test of whether the full feature set --- including sources with high $\chi^2_\mathrm{red}$ --- supports accurate classification.

\section{Feature Engineering and Classification}\label{sec:classification}

\subsection{Feature Matrix}\label{sec:features}

The feature matrix comprises 22 columns (for the PHAT filter configuration) organized into five groups (Table~\ref{tab:features}); the count varies slightly with the target galaxy's HST filter set.

\textbf{Group 1: X-ray log-fluxes (4 features).} The base-10 logarithm of the Chandra flux in the soft (0.5--1.2~keV), medium (1.2--2.0~keV), hard (2.0--7.0~keV), and broad (0.5--7.0~keV) bands.

\textbf{Group 2: X-ray hardness ratios (4 features).} Four ratios are constructed from pairwise flux combinations:
\begin{align}
    \mathrm{HR}_{SM} &= \frac{F_{x,S} - F_{x,M}}{F_{x,S} + F_{x,M}}, &
    \mathrm{HR}_{MH} &= \frac{F_{x,M} - F_{x,H}}{F_{x,M} + F_{x,H}}, 
    \label{eq:hr1} \\
    \mathrm{HR}_{SH} &= \frac{F_{x,S} - F_{x,H}}{F_{x,S} + F_{x,H}}, &
    \mathrm{HR}_{BM} &= \frac{F_{x,B} - F_{x,M}}{F_{x,B} + F_{x,M}}.
    \label{eq:hr2}
\end{align}
These ratios are distance-independent and are among the most discriminative features for X-ray source classification \citep{Prestwich2003, Yang2022}.

\textbf{Group 3: HST color indices (typically 9 features for PHAT).} Consecutive filter colors (e.g., F275W$-$F336W, F475W$-$F814W) and wide-baseline colors (e.g., F275W$-$F814W, F475W$-$F160W). For training data, SED-translated magnitudes are used, while for application data, real HST magnitudes from the Hubble Source Catalog are used.

\textbf{Group 4: X-ray to optical flux ratios (2 features).} $\log_{10}(F_{x,B}/f_{\nu,\mathrm{F475W}})$ and $\log_{10}(F_{x,B}/f_{\nu,\mathrm{F814W}})$, where $f_\nu$ is converted from the AB magnitude. This classical diagnostic separates AGN (high ratios) from stars (low ratios) and is among the most informative features \citep{Yang2022}.

\textbf{Group 5: Match quality and positional features (3 features).} The effective Chandra positional uncertainty $\sigma_\mathrm{eff}$, the normalized separation $\nu$ (Section~\ref{sec:csc_match}), and the detection significance. Including $\nu$ as a feature allows the classifier to learn that photometry from marginally matched counterparts is less reliable.

\begin{deluxetable*}{clll}
\tablecaption{Feature matrix for the XClass classifier (PHAT configuration). The 22 features are organized into five groups. The number of HST color features (Group~3) depends on the target filter set; the count shown here is for the six-filter PHAT configuration.\label{tab:features}}
\tablehead{
\colhead{Group} & \colhead{Feature} & \colhead{Description} & \colhead{Source}
}
\startdata
\multirow{4}{*}{1. X-ray fluxes}
 & $\log F_{x,S}$ & Log soft-band flux (0.5--1.2 keV) & CSC 2.1 \\
 & $\log F_{x,M}$ & Log medium-band flux (1.2--2.0 keV) & CSC 2.1 \\
 & $\log F_{x,H}$ & Log hard-band flux (2.0--7.0 keV) & CSC 2.1 \\
 & $\log F_{x,B}$ & Log broad-band flux (0.5--7.0 keV) & CSC 2.1 \\
\hline
\multirow{4}{*}{2. Hardness ratios}
 & HR$_{SM}$ & Soft--medium hardness ratio & CSC 2.1 \\
 & HR$_{MH}$ & Medium--hard hardness ratio & CSC 2.1 \\
 & HR$_{SH}$ & Soft--hard hardness ratio & CSC 2.1 \\
 & HR$_{BM}$ & Broad--medium hardness ratio & CSC 2.1 \\
\hline
\multirow{4}{*}{3. HST colors}
 & F275W$-$F336W, \dots & Consecutive filter colors & SED trans.\ / HSC \\
 & F275W$-$F814W, \dots & Wide-baseline colors & SED trans.\ / HSC \\
 & \multicolumn{2}{l}{\textit{(9 colors total for PHAT; varies by target)}} 
 & \\
\hline
\multirow{2}{*}{4. $F_X/f_\mathrm{opt}$ ratios}
 & $\log(F_{x,B}/f_{\nu,\mathrm{F475W}})$ & X-ray to optical ratio (blue) 
 & CSC + SED/HSC \\
 & $\log(F_{x,B}/f_{\nu,\mathrm{F814W}})$ & X-ray to optical ratio (red) 
 & CSC + SED/HSC \\
\hline
\multirow{3}{*}{5. Positional}
 & $\sigma_\mathrm{eff}$ & Chandra positional uncertainty & CSC 2.1 \\
 & $\nu$ & Normalized separation & Cross-match \\
 & Significance & Chandra detection significance & CSC 2.1 \\
\enddata
\end{deluxetable*}

\subsection{Data Splitting, Imputation, and Class Imbalance}\label{sec:splitting}

The training set was split via stratified random sampling into training (72\%), validation (18\%), and test (10\%) subsets, with a fixed random seed for reproducibility. Missing values (arising from incomplete survey coverage, non-detections, or failed SED translations) were imputed with the column median computed on the training set. Any remaining NaN values after imputation were filled with zero. The same procedure is applied identically to application data.

We restrict the training set to sources with at least one non-NaN predicted HST magnitude --- either from SED translation (non-SNR classes) or from direct HSC photometry (SNRs). This excludes 890 sources (7\% of the CSC-matched sample) for which SED translation failed entirely, avoiding a median-imputation artifact in which sources with all-NaN optical features cluster artificially in feature space. The final training set contains 11{,}374 sources (Table~\ref{tab:td_summary}).

To address class imbalance (Table~\ref{tab:td_summary}), the Random Forest uses \texttt{class\_weight = ``balanced\_subsample''}, recomputing class weights on each bootstrap subsample so that every tree is trained on an effectively balanced dataset. We tested SMOTE oversampling \citep{Chawla2002} as a complementary strategy but found that it provided no improvement in balanced accuracy or minority-class F1 on the optical baseline. This is consistent with minority classes already being well-separated in the restricted feature space. SMOTE is therefore not used.

\subsection{Two-Stage Random Forest Architecture}\label{sec:two_stage}

The classifier operates in two stages, motivated by the difficulty of  distinguishing LMXBs from HMXBs at optical wavelengths, where the accretion disk often dominates the donor star's spectral contribution.

\textbf{Stage~1 (broad classification)} assigns each source to one of four groups: \textbf{AGN}, \textbf{OTHER} (= LMXB + HMXB), \textbf{SNR}, and \textbf{STAR} (= LM-STAR + HM-STAR).

\textbf{Stage~2 (X-ray binary subclassification)} is trained exclusively on LMXB and HMXB sources and applied only to those that Stage~1 assigns to OTHER. The Stage~2 feature vector augments the original features with Stage~1 class probabilities:
\begin{equation}\label{eq:stage2_features}
    \mathbf{x}_2 = [\,\mathbf{x}_1 \;\|\; 
    p_\mathrm{AGN},\, p_\mathrm{OTHER},\, p_\mathrm{SNR},\, 
    p_\mathrm{STAR}\,],
\end{equation}
where $\|$ denotes concatenation. This stacked design allows Stage~2 to exploit Stage~1's calibrated uncertainty as additional discriminative signal.

\begin{figure}
\centering
\includegraphics[width=\columnwidth]{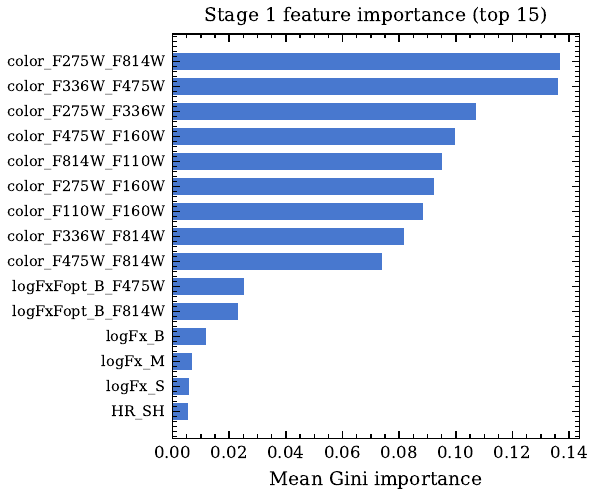}
\caption{Gini impurity-based feature importance for the top 15 features in the Stage~1 Random Forest (optical baseline), ranked by mean decrease in impurity. Error bars show the $1\sigma$ variation across the 600 trees in the ensemble. HST color features exhibit large tree-to-tree variance ($\sigma > \mu$ in some cases) because correlated colors share importance across trees depending on the bootstrap sample; X-ray features (hardness ratios, log fluxes) show tighter error bars because they are less mutually correlated.}
\label{fig:feature_importance}
\end{figure}

Both stages use the Random Forest algorithm \citep{Breiman2001} as implemented in \texttt{scikit-learn} \citep{scikit-learn}, configured with 600 trees, unbounded depth, $\sqrt{p}$ features per split, minimum 4 samples to split and 2 per leaf, and balanced subsample class weighting. An XGBoost alternative \citep{Chen2016xgb} is also implemented but is not used as the default. Figure~\ref{fig:feature_importance} shows the top 15 Stage~1 features ranked by mean decrease in Gini impurity. The UV and blue HST colors (e.g., F275W$-$F814W, F336W$-$F475W, F275W$-$F336W) carry the highest importance, followed by the X-ray-to-optical flux ratios and the broadband X-ray fluxes; the HST color features show large tree-to-tree variance because correlated colors share importance across bootstrap samples, whereas the less-correlated X-ray features have tighter error bars.

\subsection{Classification Confidence}\label{sec:confidence}

Each source receives a confidence score equal to the maximum class probability from the deciding stage. Sources with confidence $\geq 0.5$  are flagged as ``classified''; those below are flagged as ``uncertain.''

\section{Pipeline Performance}\label{sec:performance}

We evaluate classification performance using stratified 5-fold cross-validation. In each fold, a fresh Random Forest was trained on four-fifths of the data and evaluated on the held-out fifth. Class proportions are preserved via stratification and results are reported as mean $\pm$ standard deviation across folds. We report four metrics: \textit{accuracy} (fraction correct), \textit{balanced accuracy} (mean per-class recall; \citealt{Brodersen2010}), \textit{macro F1} (unweighted mean of per-class F1 scores), and the \textit{Matthews correlation coefficient} (MCC; \citealt{Matthews1975, Chicco2020}), which is robust to class imbalance. All metrics are reported both for all classifications and for the confident subset (max class probability $\geq 0.5$; Section~\ref{sec:confidence}).

\subsection{Overall Performance}\label{sec:overall_perf}

Table~\ref{tab:performance} summarizes cross-validation metrics for Stage~1 (four-class) and the full two-stage pipeline (seven-class output), with and without the confidence cut.

\begin{deluxetable}{lcccc}
\tablecaption{Cross-validation performance metrics from stratified 5-fold out-of-fold evaluation on the optical baseline (11{,}374 sources, no SMOTE). ``All'' includes every source; ``Confident'' restricts to sources with max class probability $\geq 0.5$. Values are means across 5 folds ($\pm 1\sigma$).\label{tab:performance}}
\tablehead{
\colhead{Metric} & 
\multicolumn{2}{c}{Stage 1 (4-class)} & 
\multicolumn{2}{c}{Full Pipeline (6-class)} \\
\colhead{}
& \colhead{All} & \colhead{Confident} & 
\colhead{All} & \colhead{Confident}
}
\startdata
Accuracy          & $0.999\pm0.001$ & $0.999\pm0.001$ & $0.996\pm0.001$ & $0.996\pm0.001$ \\
Balanced Accuracy & $0.994\pm0.003$ & $0.994\pm0.003$ & $0.90\pm0.02$ & $0.90\pm0.02$ \\
Macro F1          & $0.995\pm0.003$ & $0.995\pm0.003$ & $0.91\pm0.02$ & $0.91\pm0.02$ \\
MCC               & $0.999\pm0.001$ & $0.999\pm0.001$ & $0.992\pm0.002$ & $0.992\pm0.002$ \\
Completeness      & 1.000           & 1.000           & 1.000           & 1.000           \\
\enddata
\tablecomments{All sources exceed the 0.5 confidence threshold, so the 
``All'' and ``Confident'' columns are identical. The full pipeline uses 
six evaluation classes (AGN, STAR, SNR, LMXB, HMXB, CV); CV is mapped 
to OTHER by Stage~1 and is not resolved by Stage~2.}
\end{deluxetable}

\subsection{Confusion Matrices}\label{sec:confusion}

\begin{figure*}
\centering
\includegraphics[width=\textwidth]{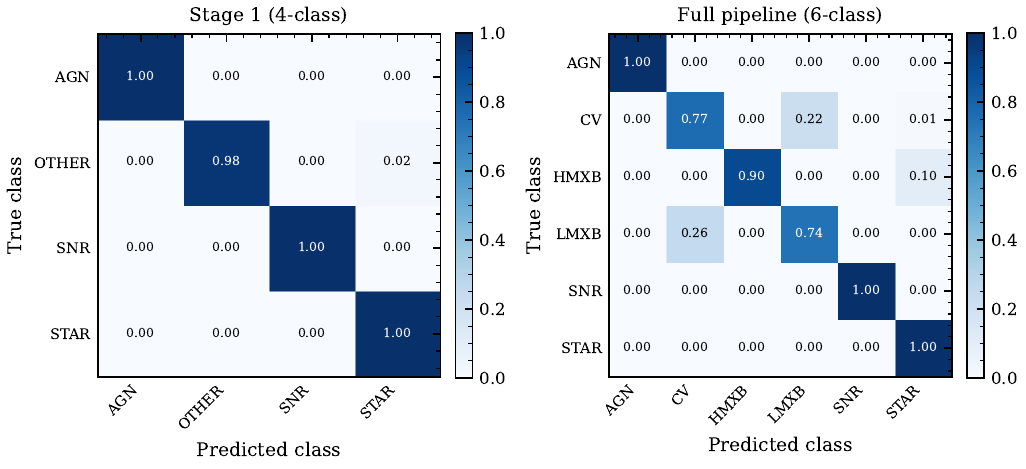}
\caption{Row-normalized (recall) confusion matrices from 5-fold 
cross-validation on the optical baseline (11{,}374 sources). \textit{Left:} 
Stage~1 (AGN, OTHER, SNR, STAR). \textit{Right:} full two-stage pipeline 
(AGN, STAR, SNR, LMXB, HMXB). Darker shading indicates higher recall; 
source counts per class are shown along the vertical axis.}
\label{fig:confusion}
\end{figure*}

Figure~\ref{fig:confusion} presents row-normalized (recall) confusion matrices for Stage~1 and the full pipeline, before and after the confidence cut.

Stage 1 achieves near-perfect separation, with the only off-diagonal entry being a 2\% OTHER$\to$STAR leakage. In the full pipeline, LMXB remains the most challenging class (F1 = 0.76), with 26\% of LMXBs (22 of 86) misclassified as CV. We find that SNR classification is now robust (F1 = 1.00), confirming that SNRs are cleanly separable when they have genuine optical photometry from the HSC rather than median-imputed features.

\begin{deluxetable}{lccc}
\tablecaption{Per-class classification performance from stratified 5-fold
cross-validation on the optical baseline (11{,}374 sources). Central values
are pooled across out-of-fold predictions; uncertainties are the fold-to-fold
standard deviation. LM-STAR and HM-STAR are merged into STAR, consistent with
the Stage~1 output; CV maps to OTHER in Stage~1 and is not resolved by Stage~2.
\label{tab:per_class}}
\tablehead{\colhead{Class} & \colhead{Precision} & \colhead{Recall} & \colhead{F1}}
\startdata
AGN  & $1.00$ & $1.00$ & $1.00$ \\
STAR & $0.999 \pm 0.001$ & $0.999 \pm 0.001$ & $0.999 \pm 0.001$ \\
SNR  & $1.00$ & $1.00$ & $1.00$ \\
HMXB & $1.00$ & $0.90 \pm 0.16$ & $0.95 \pm 0.10$ \\
LMXB & $0.77 \pm 0.11$ & $0.74 \pm 0.08$ & $0.76 \pm 0.07$ \\
\enddata
\tablecomments{Test-set sizes per fold: AGN $\sim$1046, STAR $\sim$1165,
SNR $\sim$22, LMXB $\sim$17, HMXB $\sim$8. The small XRB test sets drive the
larger fold-to-fold scatter for those classes.}
\end{deluxetable}

\subsection{Per-Class Performance}\label{sec:per_class}
\textbf{Table~\ref{tab:per_class}} shows per-class precision, recall, and F1 from the 5-fold cross-validation on the optical baseline (11{,}374 sources). Because Stage~1 maps CVs into the OTHER group and Stage~2 resolves only LMXB vs.\ HMXB, CV is never a predicted class in the full pipeline and is excluded from the figure. LM-STAR and HM-STAR are evaluated as a single STAR class, consistent with the Stage~1 output.

We find that AGN (F1 = 1.00), STAR (F1 = 1.00), and SNR (F1 = 1.00)  are all perfectly or near-perfectly classified. AGN occupy a distinct  region of high $F_X/f_\mathrm{opt}$ and relatively hard X-ray spectra \citep{Yang2022}; stars are separated by their distinctive optical/NIR colors and low X-ray-to-optical flux ratios; and SNRs are cleanly separated by their real HSC optical photometry combined with soft X-ray spectra. HMXBs achieve precision of 1.00 and recall of 0.90, yielding F1 = 0.95. LMXBs (precision 0.77, recall 0.74, F1 = 0.76) show the lowest performance: 22 of 86 LMXBs are misclassified, all of them as CV, reflecting an optical degeneracy between low-mass accreting systems that we examine in Section~\ref{sec:lmxb_cv}.
The fold-level stability is high for the populous classes: AGN, STAR, and SNR achieve F1 = 1.00 in all five folds. The minority XRB classes show larger fold-to-fold scatter, with HMXB F1 = $0.95 \pm 0.10$ and LMXB F1 = $0.76 \pm 0.07$ (per-fold mean $\pm$ standard deviation), driven by the small per-fold test sets ($\sim$8 HMXBs and $\sim$17 LMXBs); the HMXB scatter reflects test-set size rather than systematic instability.

\subsection{Confidence Threshold}\label{sec:ct_effect}

\begin{figure}
\centering
\includegraphics[width=\columnwidth]{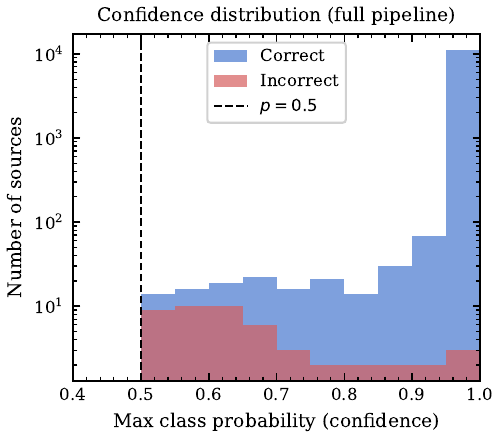}
\caption{Distribution of max class probability from 5-fold 
cross-validation on the optical baseline, shown separately for correct 
(blue) and incorrect (red) classifications on a log-scale $y$-axis. The 
dashed line marks the confidence threshold at 0.5. Incorrect 
classifications persist at intermediate confidence levels.}
\label{fig:confidence_hist}
\end{figure}

\begin{figure}
\centering
\includegraphics[width=\columnwidth]{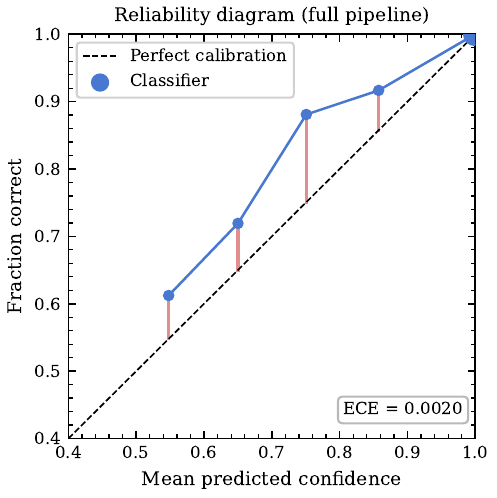}
\caption{Reliability diagram from 5-fold cross-validation on the optical baseline. The calibration curve (solid) shows the empirical accuracy as a function of predicted confidence; the diagonal represents perfect calibration. The histogram shows the number of sources per confidence bin. The expected calibration error (ECE = 0.002) indicates excellent overall calibration; the classifier is slightly underconfident across all bins.}
\label{fig:calibration}
\end{figure}

Figure~\ref{fig:confidence_hist} shows the distribution of max class probability across all cross-validated sources, separately for correct and incorrect classifications (log-scale $y$-axis). All sources exceed the 0.5 confidence threshold, with 98.2\% falling in the highest bin ($p \geq 0.9$). However, we note that incorrect classifications persist at intermediate confidence levels --- the classifier can be confidently wrong, particularly for LMXB sources with optical overlap with other compact objects. This is an inherent limitation of the Random Forest vote-fraction probability. A reliability diagram (Figure~\ref{fig:calibration}) shows that the classifier is well-calibrated overall, with an expected calibration error (ECE) of 0.002 and a Brier score of 0.003. The classifier is slightly underconfident across all bins (empirical accuracy exceeds stated confidence), which is conservative and well-suited for the intended application. The maximum calibration error (0.17) occurs in the $[0.7, 0.8)$ bin, where 40 sources reside. The full probability vector is provided for every source, enabling users to select their own thresholds.

\section{Discussion}\label{sec:discussion}

\subsection{Assumptions and Limitations}\label{sec:limitations}

The SED translation rests on several assumptions that should be taken into account when interpreting classification results.

\textbf{SED model adequacy.} The adopted models (Pickles atlas, AGN composite, two-component star+disk) capture dominant spectral characteristics but do not account for intra-class diversity. The AGN composite \citep{VandenBerk2001} does not represent the range from obscured Seyferts to LINERs, and the two-component model uses fixed donor types (K5 for LMXBs, B2 for HMXBs) whereas real donors span a range of spectral types and accretion states. These limitations are partially absorbed by the fit uncertainty ($\sigma_\mathrm{base}$), but systematic color biases may persist for atypical sources.

\textbf{UV extrapolation.} The bluest ground-based constraint (PanSTARRS $g$ at $\sim$4800~\AA) provides no coverage of the UV wavelengths probed by F275W and F336W; thus, UV predictions are model extrapolations. The 0.4~mag systematic (Section~\ref{sec:sed_uncertainty}) mitigates this, but UV features remain the least reliably translated --- particularly relevant for HMXBs, where UV excess from the massive donor is a key discriminant.

\textbf{Interstellar extinction.} The current implementation does not apply extinction corrections. \textit{Gaia} $A_V$ estimates are recorded but not yet incorporated into the translation. For Galactic training sources, variable line-of-sight extinction alters observed colors (especially in the UV), while for extragalactic targets, internal host-galaxy extinction adds further complication. Incorporating extinction --- via dereddening before SED fitting or as an additional model parameter --- is a high-priority future improvement.

\textbf{Source confusion.} In crowded galaxy fields, the Chandra 
positional uncertainty ($\sim$0\farcs5 for typical on-axis sources) can encompass multiple HST-resolved sources. The current pipeline selects the nearest HSC counterpart and includes the normalized separation $\nu$ as a feature, but this is not a substitute for full probabilistic matching. Implementing Bayesian counterpart matching \citep[e.g., NWAY;][]{Salvato2018} is planned for a future version.

\textbf{Training-set representativeness.} The training set is predominantly Galactic and may not fully represent extragalactic X-ray populations. Ultra-luminous X-ray sources and globular-cluster X-ray sources \citep[e.g.,][]{Peacock2010} are absent and would likely be misclassified as AGN, HMXB, or LMXB. Moreover, the X-ray luminosity function of the training sources reflects Galactic survey sensitivity limits, which may differ from detection thresholds at target-galaxy distances. The HSC concentration index (CI) could in principle be used to identify globular cluster counterparts in future pipeline versions.

\textbf{Flat-disk approximation.} The accretion disk is modeled as a flat ($f_\nu = \mathrm{const}$) spectrum \citep{Frank2002}, a reasonable 
zeroth-order approximation but one that ignores dependences on accretion rate, inclination, and irradiation.

\textbf{SNR optical features.} SNRs bypass SED translation and instead use real HST photometry from the Hubble Source Catalog. Of the 150 SNR training sources, 108 (72\%) have HSC detections in at least one PHAT filter; the remaining 42 --- primarily diffuse remnants in M33 below the HSC point-source detection threshold --- lack optical photometry entirely and are excluded from the training set by the $\geq 1$ optical magnitude criterion (Section~\ref{sec:splitting}). With this restriction, SNR classification achieves perfect recall and precision in cross-validation, confirming that the classifier identifies SNRs from genuine astrophysical signals rather than imputation artifacts.

\textbf{SMOTE oversampling.} We tested the classifier with and without SMOTE on the optical baseline. SMOTE provided no improvement in balanced accuracy or minority-class F1 on this dataset, consistent with the minority classes already being well-separated in the optical baseline feature space; SMOTE-generated synthetic points in this regime introduce marginal confusion rather than improving coverage. We therefore do not use SMOTE.

\textbf{Choice of classifier.} We adopt the Random Forest \citep{Breiman2001} as the default classifier rather than a gradient-boosted tree ensemble or a neural network, for several reasons specific to this application. First, the training set is small and severely imbalanced: the minority classes contain only tens of sources (Table~\ref{tab:td_summary}), with $\sim$8 HMXBs and $\sim$17 LMXBs per cross-validation fold. Random Forests are robust in this regime with minimal tuning, whereas gradient-boosted methods are more prone to overfitting on such small minority samples and require more careful regularization. Second, the Random Forest vote fraction yields well-calibrated probabilities directly, without post-hoc calibration: we measure an expected calibration error of 0.002 and a Brier score of 0.003 (Section~\ref{sec:confidence}), whereas gradient-boosted classifiers typically produce sharper but less well-calibrated scores that require Platt scaling or isotonic regression to yield reliable confidences --- a particular concern here, since the classification confidence is a primary data product. Third, the Gini importances (Figure~\ref{fig:feature_importance}) provide transparent, per-feature interpretability that supports the physical validation of the pipeline. A gradient-boosted alternative (XGBoost; \citealt{Chen2016xgb}) is implemented and selectable within the pipeline. We do not expect it to materially improve classification, since the populous classes are near-perfectly separated by the present feature set and the dominant minority-class error (LMXB$\to$CV) reflects an intrinsic feature-space degeneracy (Section~\ref{sec:lmxb_cv}) that a different tree ensemble would not resolve. We retain the Random Forest for its calibration, interpretability, and small-sample robustness, and expose XGBoost as a configurable option for future applications with larger or differently-balanced training sets.

\textbf{Gaia astrometric features.} We evaluated the inclusion of \textit{Gaia} DR3 astrometric features ($BP-RP$ and $G-RP$ colors, total proper motion, and a parallax-based foreground flag) in the classifier. We compared classifier performance with and without these features, and with the features populated from a dedicated \textit{Gaia} query of the training sources. In every configuration, the difference in classification metrics was within the fold-to-fold cross-validation uncertainty --- the \textit{Gaia} features produced no statistically resolvable change in accuracy, balanced accuracy, or any per-class F1. This is expected on physical grounds: the only discriminating signal the \textit{Gaia} astrometric features carry is Galactic parallax and proper motion, which separate nearby foreground objects from more distant ones. That discriminant does not transfer to the extragalactic application domain, where genuine target-galaxy sources --- AGN, X-ray binaries, and supernova remnants in M31, M33, and similar systems --- have no measurable \textit{Gaia} parallax or proper motion, and where the foreground stars the features would flag are already cleanly separated by their optical/NIR colors and X-ray properties. A feature whose only leverage is a Galactic distance discriminant is therefore structurally unsuited to extragalactic classification, regardless of its behavior on the predominantly Galactic training set. We accordingly exclude \textit{Gaia} astrometric features from the production feature matrix, which comprises 22 features in five groups. We retain \textit{Gaia}-based extinction estimates as a separate, promising input to the SED translation for future work (Section~\ref{sec:future}).

\textbf{Comparison with MUWCLASS.} A direct quantitative comparison with MUWCLASS \citep{Yang2022} is complicated by differences in class definitions, feature sets, and validation methodology. More fundamentally, a source-by-source comparison is ill-posed in the present context: XClass and MUWCLASS operate in disjoint feature spaces. MUWCLASS classifies Galactic sources using native ground-based photometry (PanSTARRS, \textit{Gaia}, 2MASS, \textit{WISE}), whereas XClass classifies in SED-translated HST space, with features synthesized for a specific HST filter configuration (here PHAT). The two pipelines therefore share neither a feature space nor an application domain --- MUWCLASS targets the Galactic plane, where HST coverage is sparse and filter-heterogeneous, while XClass targets HST-imaged extragalactic fields --- so there is no sample on which both can be evaluated under matched conditions. A meaningful quantitative comparison requires a common extragalactic test set with reliable ground truth, which does not yet exist. We therefore defer it to the companion paper, where validated M31/M33 classifications enable comparison with prior extragalactic work, including the X-ray-only XRB classification of \citet{Arnason2020}. We note that the two pipelines nonetheless agree at the level that is comparable: both achieve high accuracy for the populous classes, and both find confusion among the low-mass accreting systems (LMXB, CV) to be the dominant error mode (Section~\ref{sec:lmxb_cv}), indicating that this degeneracy is intrinsic to the X-ray/optical feature space rather than specific to either implementation.

\textbf{Class-aware SED model selection.} The SED translation uses the known class label to select the physical model (Table~\ref{tab:sed_models}). To quantify the impact of this choice, we performed an ablation in which all training sources were fit with a single class-blind model (Pickles templates or blackbody for all classes, regardless of label). The class-blind configurations reduced Stage~1 accuracy by 8--10 percentage points and macro F1 by 21--24 percentage points, but still achieved $>$88\% accuracy, indicating that the classifier is not wholly dependent on label-informed SED choices. We note that this design is not circular label leakage but rather the injection of physical domain knowledge: fitting an AGN with an AGN composite and a star with a stellar template produces more physically accurate translated colors than forcing a single model on all source types. Moreover, the SED translation applies only to the training set. At application time, extragalactic sources have real HST photometry directly from the HSC, so no model selection --- class-aware or otherwise --- is involved in the classification of new sources.

\subsection{Origin of the LMXB--CV Confusion}\label{sec:lmxb_cv}

The dominant classification error in the full pipeline is the
misclassification of LMXBs as CVs: 22 of 86 LMXBs (recall = 0.74) are
assigned to CV in cross-validation. We examined these failures in detail to
identify their origin. The misclassified LMXBs are not confidently wrong ---
their maximum class probabilities span 0.51--0.87 --- consistent with a
genuine feature-space degeneracy rather than a pathological model failure.

The degeneracy is, in the first instance, architectural. In the SED
translation (Section~\ref{sec:sed_models}, Table~\ref{tab:sed_models}), LMXBs
and CVs are assigned the \emph{same} two-component model: a K5 donor template
plus a flat accretion-disk spectrum. Their translated optical colors are
therefore drawn from an identical generative model and carry no
class-discriminating information: the median F475W$-$F814W is $-0.13$ for true
CVs, for misclassified LMXBs, and for correctly-classified LMXBs alike, with
negligible spread (18 of the 22 failures sit at the canonical model color).
We note that the disk-fraction parameter, which in principle could distinguish
disk-dominated from donor-dominated systems, does not break the degeneracy as
gridded: it saturates at the grid maximum ($f_\mathrm{disk} = 0.5$) for the
large majority of both LMXBs and CVs, and does not differ significantly between
correctly- and incorrectly-classified LMXBs (Mann--Whitney $p = 0.09$).

With the optical colors degenerate, the LMXB/CV decision rests almost entirely
on the X-ray features. The LMXBs that fail are precisely those whose X-ray
properties overlap the CV distribution: their X-ray-to-optical flux ratios are
statistically indistinguishable from true CVs ($p = 0.94$) while differing from
correctly-classified LMXBs ($p = 0.02$), and they are systematically X-ray
fainter (median detection significance 22 versus 65 for correctly-classified
LMXBs; $p = 0.001$). The X-ray hardness ratios do not separate the groups
($p > 0.6$), so the confusion is driven by the X-ray-to-optical normalization
and source faintness rather than by spectral shape. Notably, SED-fit quality
($\chi^2_\mathrm{red}$), the number of photometric bands, and cross-match
quality ($\nu$) do \emph{not} differ between the failed and correctly-classified
LMXBs ($p > 0.15$ in all cases), ruling out poor data as the primary cause.

In summary, the LMXB$\to$CV confusion arises from a shared SED model that
renders the two classes optically identical, leaving classification to the
X-ray features; the LMXBs that fail are the X-ray-faint subset whose
X-ray-to-optical ratios coincide with the CV population. This is a genuine
astrophysical and methodological degeneracy --- low-mass accreting systems with
similar optical SEDs and overlapping X-ray properties --- rather than a tuning
artifact. Class-specific donor templates spanning a range of spectral types,
or the inclusion of X-ray variability and timing information
(Section~\ref{sec:future}), would be the most direct routes to breaking it.

\subsection{Future Improvements}\label{sec:future}

Several enhancements are planned. Incorporating \textit{Gaia} extinction estimates, and 3D dust maps \citep{Green2019} into the SED fitting would improve translated colors, particularly for training sources in the Galactic plane. Expanding the training set --- especially the underrepresented LMXB, HMXB, CV, and SNR classes, and potentially incorporating validated extragalactic classifications from M31/M33 as feedback --- would directly improve minority-class performance. In addition, adding X-ray variability features from CSC~2.1 would enhance discrimination between transient and persistent sources, and implementing full Bayesian counterpart matching via NWAY \citep{Salvato2018} for both training and application would improve robustness in crowded fields. As the training set grows, deep learning architectures may offer improved feature interactions, the modular pipeline design allows the classifier to be swapped without modifying upstream processing. Finally, the SED translation framework can be adapted to future facilities: \textit{Athena} \citep{Nandra2013}, AXIS \citep{Mushotzky2018}, and the Einstein Probe \citep{Yuan2022} will detect large numbers of extragalactic X-ray sources, while the Vera C.~Rubin Observatory \citep{Ivezic2019}, the \textit{Roman Space Telescope}, and \textit{Euclid} \citep{Laureijs2011} will provide deep optical/NIR coverage. Adapting XClass to use Rubin/LSST $ugrizy$ as the application filter system would enable classification beyond HST's limited field of view.


\section{Summary and Conclusions}\label{sec:summary}

We have presented XClass (X-ray Classifier for extragalactic sources), an end-to-end machine-learning pipeline for classifying extragalactic X-ray point sources into seven astrophysical classes (AGN, LMXBs, HMXBs, CVs, LM-STARs, HM-STARs, SNRs). The main results are as follows.

\begin{enumerate}

\item We developed an SED translation procedure that bridges the photometric gap between ground-based training data (PanSTARRS and 2MASS) and HST application data by fitting class-appropriate physical spectral models and convolving them through HST filter curves, producing synthetic magnitudes in a common feature space. The approach is general and applicable to any HST filter configuration. SED fit quality varies across the training set, but the uncertainty propagation automatically down-weights poorly constrained translated magnitudes. The cross-validation metrics in Section~\ref{sec:performance} represent the end-to-end performance of the full pipeline (Section~\ref{sec:sed_translation}).

\item We assembled a training dataset from ten VizieR catalogs of Galactic X-ray sources and extragalactic SNR catalogs (M33, NGC~6946, M83), cross-matched to CSC~2.1 detections with photometry from three ground-based surveys (see Table~\ref{tab:td_summary}).

\item We constructed a 22-feature matrix combining X-ray fluxes and hardness ratios, SED-translated HST colors, X-ray--to--optical flux ratios, and positional match-quality metrics (Section~\ref{sec:features}).

\item We implemented an asymmetric two-stage Random Forest: Stage~1 separates AGN, OTHER (LMXB+HMXB), SNR, and STAR; Stage~2 resolves OTHER into LMXB or HMXB using an augmented feature vector that includes Stage~1 probabilities (Section~\ref{sec:two_stage}).

\item We evaluated performance via stratified 5-fold cross-validation on the optical baseline (11{,}374 sources, no SMOTE). Stage~1 achieves 99.9\% accuracy; the full pipeline achieves 99.6\% accuracy and balanced accuracy of 0.90 (macro F1 = 0.91, ECE = 0.002). AGN, STAR, SNR, and HMXB are classified with F1 $\geq 0.95$; LMXB confusion (F1 = 0.76) is the dominant remaining source of error (Section~\ref{sec:performance}).

\item The pipeline is implemented as a modular, open-source Python package with centralized configuration, per-source disk caching, and a CLI supporting stage-by-stage execution.

\end{enumerate}

XClass will be applied to M31 and M33 in a forthcoming companion paper. The SED translation framework is adaptable to future facilities, including \textit{Athena}, AXIS, the Einstein Probe, the Vera C.~Rubin Observatory, and the \textit{Roman Space Telescope}, supporting classification of ever-larger extragalactic X-ray source samples.

\begin{acknowledgments}
We thank the anonymous referee for a careful and constructive report that improved the clarity and rigor of this paper. Support for this work was provided by the National Aeronautics and Space Administration (NASA) through Chandra Award Number AR8-19009A issued by the Chandra X-ray Center, which is operated by the Smithsonian Astrophysical Observatory for and on behalf of NASA under contract NAS8-03060, and by NASA under award number 80NSSC22K1119 issued through the Astrophysics Data Analysis Program (ADAP). This research has made use of data obtained from the Chandra Source Catalog, provided by the Chandra X-ray Center (CXC) as part of the Chandra Data Archive. Claude \citep{Anthropic2025} was used for assistance with code editing, data visualization, and language refinement. The XClass pipeline, the machine-readable training catalog, and the trained classifier models are publicly available at \doi{10.5281/zenodo.20838124}.
\end{acknowledgments}

The HST data used in this paper can be found in MAST: \dataset[10.17909/T97P46]{http://dx.doi.org/10.17909/T97P46} (Hubble Source Catalog) and \dataset[10.17909/s0zg-jx37]{http://dx.doi.org/10.17909/s0zg-jx37} (PanSTARRS DR2).

\facilities{CXO, HST(ACS), HST(WFC3), MAST}

\software{astropy \citep{astropy2013,astropy2018,astropy2022},
          scikit-learn \citep{scikit-learn},
          numpy \citep{numpy},
          matplotlib \citep{matplotlib},
          pandas \citep{pandas},
          Claude \citep{Anthropic2025},
          XClass \citep{XClass2026}
         }

\bibliography{bibliography}{}
\bibliographystyle{aasjournalv7}

\end{document}